\documentclass[twocolumn]{aa}
\usepackage{graphicx}
\usepackage[intlimits,sumlimits]{amsmath}
\usepackage{deluxetable}
\usepackage{times,epsfig} 
\usepackage{natbib}
\usepackage{amssymb}
\usepackage{mathrsfs}  
\bibpunct{(}{)}{;}{a}{}{,} 

\usepackage{amsmath}
\usepackage[english]{babel}
\usepackage{longtable}
\usepackage{url}
\usepackage{hyperref}

\newcommand{\beqa}{\begin{eqnarray}} 
\newcommand{\eeqa}{\end{eqnarray}}

\newcommand{\bsub}{\begin{subequations}}
\newcommand{\esub}{\end{subequations}}
\newcommand{\beal}{\begin{align}}
\newcommand{\ealn}{\end{align}}

\begin{document}

\title{Do Wolf-Rayet stars have similar locations in hosts as type Ib/c 
supernovae  and long gamma-ray bursts?}
\titlerunning{WR star locations in their host galaxies}

\author{
G.~Leloudas\inst{1}
\and J.~Sollerman\inst{1,2} 
\and A.~J.~Levan\inst{3}
\and J.~P.~U.~Fynbo\inst{1}
\and D.~Malesani\inst{1}
\and J.~R.~Maund\inst{1,4,5}
}

\institute{Dark Cosmology Centre, Niels Bohr Institute, University of Copenhagen, Juliane Maries Vej 30, 
2100 Copenhagen \O, Denmark \email{giorgos@dark-cosmology.dk}
\and The Oskar Klein Centre, Department of Astronomy, Stockholm University, AlbaNova, 10691 Stockholm, Sweden
\and Department of Physics, University of Warwick, Coventry CV4 7AL, UK
\and Department of Astronomy \& Astrophysics, University of California, Santa Cruz, CA 95064, USA
\and Sophie \& Tycho Brahe Fellow
}

\date{Received 27 November 2009 / Accepted 31 March 2010}

\abstract{} 
{
We study the distribution of Wolf-Rayet (WR) stars and their subtypes with 
respect to their host galaxy light distribution.
We thus want to investigate whether WR stars are potential progenitors 
of stripped-envelope core-collapse supernovae (SNe) and/or 
long-duration gamma-ray bursts (LGRBs). 
}
{
We derived the relative surface brightness (fractional flux) at the locations of WR stars and compared with similar results for LGRBs and SNe.
We examined two nearby galaxies, M~83 and NGC~1313,
for which a comprehensive study of the WR population exists.
These two galaxies contain a sufficiently large number of WR stars 
and sample  different metallicities.
To enable the comparison, the images of the galaxies were
processed to make them appear as
they would look at a higher redshift.
The robustness of our results against several sources of uncertainty was 
investigated with the aid of Monte Carlo simulations. 
} 
{
We find that the WC star distribution favours brighter
pixels than the WN star population. 
WC stars are more likely drawn from the same distribution as SNe~Ic 
than from other SN distributions,
while WN stars show a higher degree of association with
SNe~Ib. 
It can also not be excluded that WR (especially WC) stars are related to LGRBs.
Some differences between the two galaxies do
exist, especially in the subtype distributions, and may  
stem from differences in metallicity.
}
{
Although a conclusive answer is not possible,
the expectation that WR stars are 
the progenitors of SNe~Ib/c and LGRBs survives this test. 
The trend observed between the distributions of WN and WC stars, 
as compared to those of SNe~Ib and Ic, is consistent
with the theoretical picture that SNe~Ic result from progenitors that have been stripped of a larger part of their envelope.
}


\maketitle

\section{Introduction}

To understand the origin of cosmic explosions like supernovae (SNe) and 
gamma-ray bursts (GRBs), 
it is important to study and constrain their environments.
\citeauthor{F06} (\citeyear{F06}; hereafter F06) presents a sample 
of 32  long-duration GRB (LGRB) host galaxies 
and has developed a new technique to show that  LGRBs have a
tendency to occur in the brightest pixels of their host galaxies.  
This contrasts to a comparison sample of core-collapse (CC)
SNe where the SN locations instead follow the light distribution of their hosts. 
\citeauthor{K08} (\citeyear{K08}; hereafter K08)  
further shows that not all CC SNe follow the same host galaxy
light distribution, with SNe Ic strongly skewed towards the
brightest regions of their hosts.  
The SN~Ic population is thus broadly consistent with that of
LGRBs. Indeed, SNe~Ic, and in particular broad-lined SNe~Ic, are
so far the only type that have been been firmly observationally connected to
LGRBs \citep{galama98,hjorth2003,stanek,malesani2003lw} or their lower
energy siblings, X-ray flashes \citep{pian2006aj,campana2006aj,sollerman2006aj}.

Using a similar method, \cite{AndersonJames} 
mapped the association of nearby SNe with H${\alpha}$ regions.
They find that SNe~Ic 
trace the H${\alpha}$ emission, and recent star formation, 
to a higher degree than other CC SN types.
This was attributed to SNe~Ic 
having more massive progenitors 
than SNe~Ib, which are in turn more massive than SNe~II.  
\cite{Larsson2007} modeled the F06 results and derived
minimum masses for CC SNe and LGRBs of 8 and 20~$M_{\sun}$,
respectively.  A theoretical-modeling approach was also taken by
\cite{raskin08}, who uses the F06 method and a simulated
solar-metallicity spiral galaxy to predict the (increasing) minimum
cut-off masses of the progenitors of SNe~II and Ic.

The next step is to expand this type of study to the potential progenitors of
these cosmic explosions. 
The most plausible candidates are Wolf-Rayet (WR) stars
\citep{woosleybloom,WRreview}.
In single-star evolutionary models, 
WR stars are the final phases in the life of very
massive stars \citep[minimum initial mass $>$22--37 $M_{\sun}$, 
depending on the metallicity and rotation;][]{MeyMae2005}, 
which have shed their hydrogen envelope. 
WR stars can also result from close binary star interaction, 
through Roche-lobe overflow, 
in which case the minimum mass can be decreased down to 
$\sim$15~$M_{\sun}$\ \citep{Eldridge2008}. 
A comprehensive review of WR stars is
given by \cite{WRreview}. Here we restrict ourselves to giving the
background information that is essential for the purposes of this
study: WR stars are further divided into nitrogen-rich (WN) and
carbon-rich (WC) stars.  
WN and WC stars are believed to give rise to
supernova explosions of Type~Ib and~Ic, 
respectively \citep{WRreview,Georgy2009}.
Depending on their emission line properties, width, and appearance,
they can be further divided into
`early' or `late' types (WNE, WNL, WCE, and WCL, respectively).
The WR star populations strongly
depends on metallicity, with both the total number of WR stars and the
relative ratio of WC/WN stars increasing dramatically with 
metallicity \citep{WRreview,MeyMae2005}. This is attributed to the dependence
of winds on metallicity, which leads to much more effective mass loss in
the presence of metals \citep{vink}.

Our purpose here is to study the distribution of WR stars and their subtypes 
within their hosts, using the method applied by F06 and K08. 
The main motivation is the following: if WR stars are the immediate 
progenitors of SNe~Ib/c and LGRBs, then we would also expect their 
distributions with respect to the host galaxy light to be similar. 
Since the progenitors of SNe Ib/c still evade direct detection 
\citep[see e.g.][]{maund04gt,ckockett07gr,smarttARAA}, 
this method can give us hints to their nature.
The same is true for LGRBs, for which a direct progenitor detection 
seems impossible, at least in the near future.
It is noteworthy that WR features have been clearly identified in the 
host of the most nearby LGRB (GRB~980425/SN~1998bw), 
albeit with a considerable offset to the explosion position \citep{hammer06,christensen08}
and more recently in four more LGRB hosts \citep{2010arXiv1001.2476H}.

This paper is structured as follows. In Sect.~\ref{sample} 
we discuss the galaxies 
chosen for studying the
WR distribution. In Sect.~\ref{methods} 
we outline the methods used throughout the paper, and in Sect.~\ref{results} 
we present our results. Section~\ref{disc} contains a discussion of the results and several uncertainty factors,
and Sect.~\ref{conc} summarizes our conclusions.

\section{Galaxy selection}
\label{sample}

It is essential for our purposes to use galaxies where the WR star 
population has been mapped accurately and systematically 
in an unbiased way, as much as that is possible.
For this reason we used the results of Crowther and collaborators, 
who have recently identified and spectroscopically classified the 
WR stars in a number of galaxies beyond the Local Group, in an essentially 
complete way \citep{NGC300_I,M83,NGC1313}. 
The method used is explained in detail
in, e.g., \cite{NGC1313} and includes the following steps:
first imaging of the galaxy is obtained and then regions with
candidate WR stars are identified by their excess in a narrow-band
\ion{He}{ii} $\lambda$4684 filter over the continuum. Subsequently,
the regions containing WR stars are confirmed with spectroscopy of
the candidates. Their spectral type (WN, WC, early, late) is determined, 
and the number of WR stars is estimated by fitting template spectra 
to the flux-calibrated, integrated spectrum of the WR-star region.
By excluding galaxies where the survey did not cover the entire 
galaxy \citep{NGC300_I,NGC3125}, we decided to rely on the following two 
galaxies: M~83 \citep{M83} and NGC~1313 \citep{NGC1313}.

We did not use the existing catalogues for other (very nearby)
galaxies, such as the LMC \citep{LMCcatalogue}, the SMC
\citep{SMCcatalogue}, or other Local Group galaxies 
\citep[][]{masseyM33}, 
because they suffer from
several incompleteness-related issues (chance discoveries, Malmquist
bias) and have complicated revision histories \citep[see
e.g.][]{masseyM33}. Another practical disadvantage with these
galaxies is their large angular size.  The WR galaxy catalogue in
\cite{WRgalaxySchaerer} is also unsuitable for our purposes, because it is 
a list of diverse objects that have been defined from the appearance of
a broad \ion{He}{ii} feature in their integrated spectra.

\begin{figure}
\includegraphics[width=\columnwidth,clip=]{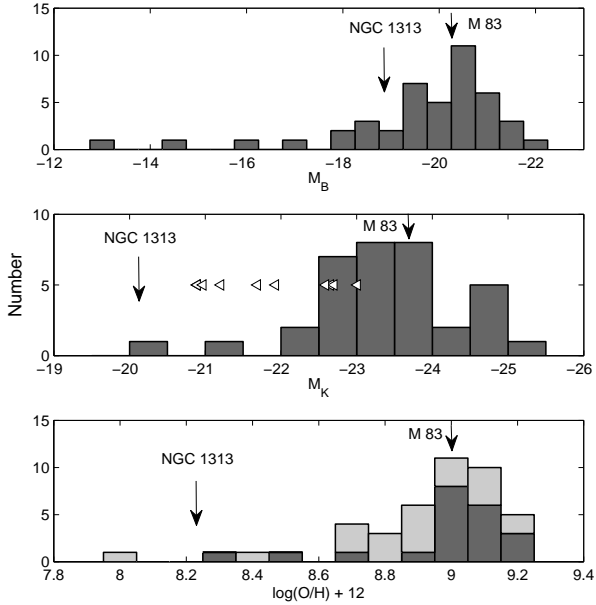}
\caption{Global properties of M~83 and NGC~1313 compared to those of the K08 SNe Ib/c hosts.
Histograms of the absolute magnitudes $M_B$, $M_K$ (a better tracer of the total stellar mass), and metallicity are presented.
The information presented here was collected from NED, SDSS, and from \cite{Prieto2008}, for the metallicities. 
These are global metallicities on the scale used by \cite{2004ApJ...613..898T}.
The $M_K$ histogram is incomplete towards the faintest galaxies 
and upper limits, compatible with the 2MASS survey upper limit, are provided (white arrows).
To get an indicative value for the galaxies with no metallicity reported in  \cite{Prieto2008}, i.e. the ones not included in SDSS DR4, 
we used an approximate luminosity-metallicity relation (see Sect.~\ref{metal}). These galaxies are represented by the light-grey shaded histogram.
The corresponding values for M~83 are marked, indicating that this galaxy is very typical of the K08 sample, just closer.
NGC~1313, on the other hand, is a galaxy that is fainter and more metal-poor, hence more reminiscent of LGRB hosts.
}
\label{fig:histo}
\end{figure}

The number of WR stars contained in
the two selected galaxies ($\sim$1000 and 100, respectively)
exceeds the number of 32 LGRB used by F06
and 44 SNe Ib/c used by K08 (out of 504 SNe of all types).  
While F06 and K08 constructed their distributions by looking at one
explosion site per galaxy for many galaxies, 
discovered in searches that are not 
unbiased themselves,
we look at many (potential) explosion sites in a few galaxies.  
Such a comparison should be valid, 
as long as the galaxies we choose are not 
different from the typical galaxies studied by K08 and F06.  While
this statement is obviously more problematic for the high-redshift F06
hosts (see also discussion in K08), 
we can compare the
global properties of M~83 and NGC~1313 (see Sects.~\ref{subsecM83} and \ref{subsecNGC1313})
to those of the K08 SNe Ib/c
host sample. 
M~83 turns out to be a galaxy very typical of
the K08 sample  in terms of absolute magnitude, metallicity
(Fig.~\ref{fig:histo}), and morphological type - with its only difference
that it is  closer to us. NGC~1313 is somewhat different, although
such galaxies are present in the K08 sample: 
fainter, metal-poor, and more irregular, it is more
reminiscent of the LGRB hosts of F06. 
It is important for our study, as we will see, that the two galaxies probe two
different metallicity limits.

Below, a more detailed description of the individual galaxies used in this study is given,
containing only those details necessary for our discussion. 
Some of their key properties are also summarized in Table~\ref{tab:galprop}.

\begin{table*}
\caption{Key properties of M~83 and NGC~1313.$^a$}
\label{tab:galprop}
\centering
\begin{tabular}{lcccccccccccc}
\hline\hline
Galaxy &Metallicity & $D$ &	 Pix. scale$^b$ &$M_B$ &$M_K$& Morph. & Conf. reg.$^c$ &WN$^d$ & WC$^d$ &Cand. reg.$^e$\\
&\tiny{($\log$(O/H)+12)} &(Mpc)   &(pc$^2$ pix$^{-1}$) & (mag) & (mag) & & &WNE$+$WNL & WCE$+$WCL & \\
\hline
M~83 & 9.0 & 4.5 &  5.3$^2$ & $-$20.2 & $-$23.7 & SAB(s)c &  132  & 232$+$250 & 28$+$526 & 89   \\
NGC~1313 & 8.23 & 4.1  & 4.0$^2$ & $-$18.9 & $-$20.1 & SB(s)d  & 70  & 29$+$22 & 34$+$0 & 12$^f$ \\
\hline
\end{tabular} \\
\begin{tabular}{lll}
$^a$ Compiled from \cite{M83,NGC1313} and NED.\\
$^b$ In the images from ESO/MPI 2.2m (+WFI) and VLT/FORS used in the analysis. A galaxy at $z=0.01$ in the SDSS 2.5m images used by K08 \\
would have $\sim$80$^2$ pc$^2$ pix$^{-1}$.\\
$^c$ Confirmed regions containing one or more WR stars.\\
$^d$ Total number of stars per subtype used in this paper. 
These numbers may differ slightly from those reported
in the literature, for the following\\ reasons: WN5-6 stars have been included in the WNE distribution; WO in WCE; transitional WN/WC have been included in both\\
WN and WC. The total number of WR stars is obtained by adding their corresponding subtypes.\\
$^e$ Candidate regions for which spectroscopy has not been conducted.\\
$^f$ 11 of these are photometrically consistent with WN stars.\\
\end{tabular}
\end{table*}

\subsection{M~83}
\label{subsecM83}

M83 
is a nearby (4.5 $\pm$ 0.3 Mpc), Milky-Way type SAB(s)c spiral galaxy 
with a super-solar
metallicity\footnote{Metallicities in this paper are expressed as oxygen abundances $\log ($O/H$) + 12$.} of 9.0 \citep[][and references therein]{M83}.
In total, 132 WR regions containing $\sim$1000 $\pm$ 300 WR stars were
spectroscopically confirmed by \cite{M83}, while 89 more regions are labeled as
candidates, still awaiting spectroscopic follow-up.  
As part of the same study, \cite{M83_WC} points out the
large number of WCL stars found in M~83, 
directly related to its high metallicity.  

\cite{M83} were not able to look for WR stars in the nucleus of M~83 
(owing to saturation in the images).  
However, K08 also provide results after the removal of the bulge, 
for galaxies with a significant bulge contribution, so comparison is still possible.
For our analysis we used a wide field image obtained using the ESO/MPI 2.2m (+WFI).

While very different from the LGRB host galaxies (F06),
M~83 is a prodigious SN producer, with 6 observed SNe in the 20th century.  
Although most of them remain unclassified, 
one of them is a prototypical SN Ib \citep[SN~1983N;][]{uomoto85,Elias1985}.
The host site of SN 1983N is not associated with a confirmed WR site and 
is $\sim 7 \arcsec$ away from the nearest candidate region.
If an isolated WR star (such as several others
identified in M~83) had been responsible for the 1983 explosion, it would
of course have disappeared \citep[e.g.][]{MaundDisap} 
from the frames obtained later by \cite{M83}.  

\subsection{NGC 1313}
\label{subsecNGC1313}

NGC~1313 is a SB(s)d spiral at a distance of 4.1 $\pm$ 0.1 Mpc 
and has a metallicity of 8.23, i.e., 
intermediate between the SMC and the LMC 
\citep[][and references therein]{NGC1313}. 
This galaxy is more reminiscent of 
the LGRB hosts than M~83, both regarding the more irregular shapes of
the LGRB hosts (F06)
and their, usually, low metallicity \citep{sollerman05,modjaz,Savaglio2009}.
Two Type II SNe have been recorded in this galaxy, SNe 1962M and 1978K.
\cite{NGC1313} report on the spectroscopic confirmation of 70 WR
regions (success rate of 85\% over the candidates followed with
spectroscopy), while 12 more regions remain candidates and are
photometrically consistent with WN stars. Unlike M~83, few of the
identified regions contain more than one WR star, with their total number
estimated to be between 84 and 115.  

Our analysis was performed on the 
VLT/FORS images obtained by
\cite{NGC1313}. We included the single WO star 
(star \#31)\footnote{
We follow the
numbering of \cite{M83} and \cite{NGC1313}.}  in the WC stars, while
the transitional WN/C star \#11 was included in both the WN and WC
distributions. For the WN5-6 stars we adopted a WNE classification,
although this choice is not unique \citep{WRreview}.
The results presented in this paper do not change, however, if we include them in the WNL distribution instead.

\section{Methods}
\label{methods}

\subsection{Fractional flux}

As fractional flux of a pixel belonging to a galaxy, we define the sum
of all counts in pixels less bright than the pixel in question, over
the sum of all counts in all the pixels belonging to the galaxy. This
is the same definition as the one used by F06 and K08.

\subsection{Pixel detection}

As in F06 and K08, the SExtractor software \citep{SEx} was used to
identify the pixels that belong to the galaxies.  The parameters used
were similar to the ones in these references to make the comparison
as close as possible. We used $B$-band images, 
which correspond to the same (rest-frame) wavelength window examined by F06 and K08.

\subsection{Pre-processing of images}
\label{sec:pre-proc}

Identification of individual stars in other galaxies requires that
the galaxies are nearby. As a consequence, the apparent dimensions of
M~83 and NGC~1313 (at $z=$ 0.0017 and 0.0016, respectively) are much
larger than the SN hosts of K08,
 and their images cover 
thousands of pixels on the CCD.  
While usually an advantage, for our type of analysis this can pose problems
in two different ways.  First, a considerable number of foreground
stars are superimposed on the image and contribute to the pixel
fractional fluxes.  It is desirable to remove the
foreground star 
point-spread functions (PSF), but  
it is not straightforward to identify which stars are in the
foreground and which belong to the host galaxy. It has been shown
that it is possible to distinguish between massive stars in other
galaxies and foreground dwarfs by use of colour-colour diagrams
\citep{RSGlocal,RSGinM31} or other kinematical techniques involving
spectroscopy \citep{YSGinM31}.  However, it was not necessary to
resort to such detailed techniques 
for our purposes, 
because it is sufficient to remove only the brightest stars that
are clearly in the foreground and contribute with significant light.
To this end, all stars brighter than 
$\sim$20 mag were removed
from our images by subtracting their PSF
with tasks in the package
{\tt daophot} in IRAF\footnote{IRAF is distributed by the National
Optical Astronomy Observatory: \url{http://iraf.noao.edu/iraf/web/}.}.
At the distances of our galaxies, there is no degeneracy below this
limit, and all stars can safely be considered Galactic.  To assess the
effect of stars fainter than 20 mag, we performed a Monte Carlo (MC)
simulation:  the expected number of foreground stars in the field
\citep{BahcallSoneira} and their corresponding counts,  per magnitude
bin, were removed randomly and repetitively from our images. This
experiment showed that our results are not sensitive to their
presence.

\begin{figure}
\includegraphics[width=\columnwidth,clip=]{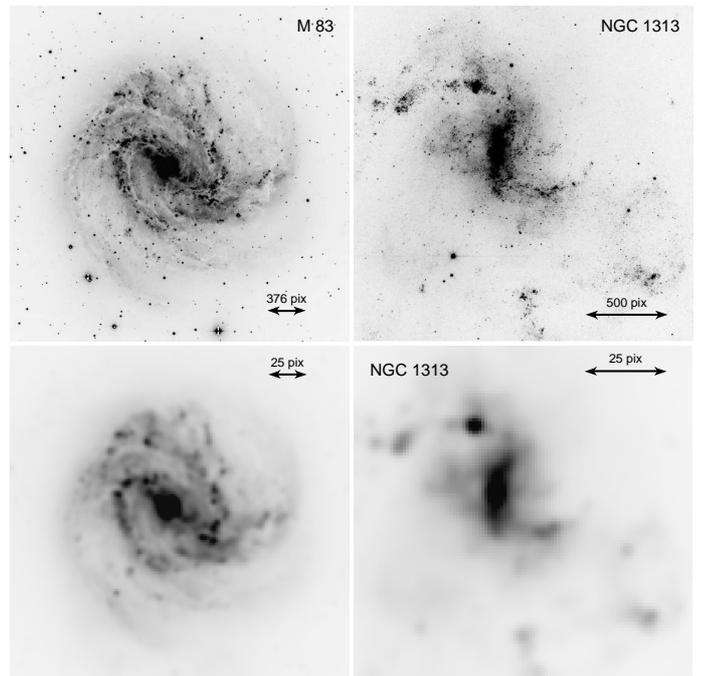}
\caption{{\bf Top left:}
ESO/MPI 2.2m (+WFI) $B$-band image of M~83 ($z=0.0017$, $D=4.5$ Mpc). 
The displayed part is 3300$\times$3300 pixels. 
{\bf Top right:} VLT/FORS $B$-band image of NGC~1313 ($z=0.0016$, $D=4.1$ Mpc). The image is 2048$\times$2048 pixels. 
{\bf Bottom:} Processed images, which simulate how the galaxies above would 
look at $z=0.01$ at the SDSS 2.5m telescope.
Bright foreground stars have been removed and the images have been binned 
such that one pixel side is $\sim$80 pc along.
A Gaussian filter has been used to smooth the resulting images to an 
FWHM of $\sim$3.2 pixels, typical of the SDSS frames.
The pixel dimensions of the resulting images are 
220$\times$220 and 108$\times$108, respectively. 
In each panel, a double arrow spanning 2 kpc across has been drawn and the corresponding pixel scale is noted.
}
\label{fig:galimages}
\end{figure}

Second, each of our pixels contains light from a much smaller 
physical area
than the ones studied by K08:  the pixel scale in our images is
$\sim$4 and 5 pc (along a pixel side) for NGC~1313 and M~83,
respectively.  On the other hand, the SN Ib/c hosts of K08 have a
median redshift of $z\gtrsim0.01$, which corresponds to a distance of 
$\sim42$ Mpc for $H{_0}=72$~km~s$^{-1}$~Mpc$^{-1}$.  
At this distance,
one pixel on the SDSS 2.5m telescope, used by K08, corresponds to an
area of $\sim 80 \times 80$~pc$^2$.  
Consequently the pixel fractional fluxes 
they quote refer to these large areas.  
To enable the comparison
we thus binned our images by
factors of 15-19 in order to `bring our galaxies' to a redshift of
$z=0.01$.  Subsequently, the images were convolved with a Gaussian in
order to simulate the typical seeing of 3.2 pixels measured on the K08 SDSS
plates.
The final images are shown in Fig.~\ref{fig:galimages}.

\subsection{The metallicity dependence}
\label{metal}

As mentioned above, both the total number of WR stars and the WC/WN
ratio increase drastically with metallicity.  This trend is also
observed for M~83 and NGC~1313 \citep{M83,NGC1313}.  If WR stars are
the progenitors of SNe Ib/c, we therefore expect to observe relatively
more SNe Ib/c in galaxies with high than low metallicity.  This has, in
fact, been shown by  \cite{PB03}, \cite{Prieto2008} and \cite{BP09}.
\cite{AndJames2009} even claim that they can disentangle the metallicity
dependence  
from the mass dependence and that SNe Ic come from higher metallicity
environments than SNe Ib.  
On the other hand, LGRBs are often found
in low-metallicity environments 
\citep{sollerman05,stanekMetals,Savaglio2009,levesque10grbs}, 
especially in comparison to  broad-lined SNe Ic that are not associated with known LGRBs \citep{modjaz}.
However, it has not been completely settled whether all LGRB hosts are metal poor or if we are simply missing
more metal-rich hosts owing to a dust-obscuration bias \citep[e.g.][]{fynbo2009,Lev020819,svensson}. 
Because of these strong metallicity dependencies, in the following section, our results
will be presented separately for the metal-rich M~83 and the metal-poor NGC~1313.

We have also attempted to look for possible metallicity signatures in
the K08 sample. To do so, the K08 SN Type Ib and Ic fractional flux
distributions were divided into two equal number bins of `high' and
`low' metallicity.  
For the host galaxies for which \cite{Prieto2008} do not report
metallicities,  we followed
\cite{PB03} in using the galaxy global luminosity as a proxy for
metallicity.  
Although it has been shown that the local metallicities
at the explosion sites can differ from the global metallicity or its
proxies \citep[e.g.][]{modjaz,CT2770}, this should not
pose any problems for the rough separation of the sample in two bins.  
It was thus assumed that the galaxies without reported metallicities have
metallicities consistent with the best fit luminosity-metallicity
relation, as derived from the galaxies with reported metallicities
(light-grey histogram in Fig.~\ref{fig:histo}, lower panel).
We see no convincing evidence that the high and low metallicity SNe are
different in terms of their fractional flux distributions.  
This investigation is, however, clearly limited by the small number
of SNe in the sample and by the absence of very low (i.e. subsolar) 
metallicities.

\section{Results}
\label{results}

To compare with the fractional flux distributions of LGRBs
(F06) and various types of SNe (K08), it is necessary to 
consider the number of WR stars contained within each WR confirmed
region, because each one of them is a potential progenitor of a SN
and/or LGRB explosion.  \cite{M83} and \cite{NGC1313} give estimates
for the numbers or WR stars contained in each region, as well as for
the corresponding errors.  The number of stars per region is simply
taken into account by, e.g., including the fractional flux value of the
pixel that hosts 5 WR stars 5 times in the corresponding
distribution. 

\begin{table}
\caption{ Fractional fluxes at the locations of the WR stars.$^a$}
\label{tab:FFpart}
\centering
\begin{tabular}{lrrrr}
\hline\hline
& ID$^b$ & Subtype$^c$  &   Number & FF$^d$ 	\\
  \hline
 &   1   &    WNL   & 	   3  &   0.29    \\  
  &  2   &    WCL   & 	   2  &   0.34    \\  
  &  3   &    WCL   & 	   2  &   0.26    \\  
{\bf M~83} &    4   &    WNE   & 	  16  &   0.48    \\ 
&    5   &    WCL   & 	   8  &   0.48    \\  
 &   6   &    WCE   & 	   3  &   0.06    \\  
&   \nodata &   \nodata &	\nodata &  \nodata   \\ 
&  133   &    cand.   &   \nodata  &   0.25      \\  
&  134   &    cand.   &   \nodata  &   0.54      \\  
  \hline
&    1   &    WNE   & 	   1  &   0.37      \\  
 &   2   &    WCE   & 	   1  &   0.30      \\  
{\bf NGC~1313}&   3   &    WNL   & 	   1  &   0.06      \\
&    \nodata &   \nodata &	\nodata &  \nodata   \\ 
&   28   &    cand.   &   \nodata  &   0.38      \\  
\hline
\end{tabular} \\
\begin{tabular}{lll}
$^a$ An indicative portion of this table is shown. For the full table refer \\ to the Appendix.\\
$^b$ We follow the exact numbering of \cite{M83} and\\ \cite{NGC1313}. Additional info (e.g. fluxes, errors in\\ star numbers) can be found in these references.\\
$^c$ `cand.'~stands for candidate region.\\
$^d$ These fractional fluxes refer to a binning of 15 and 19 for the\\ images of M~83 and NGC~1313 we have used, respectively. \\
Although we have shown that the results do not change significantly\\ with the choice of binning, individual fractional fluxes can vary.\\
\end{tabular}
\end{table}

\begin{figure*}
\includegraphics[width=\textwidth,clip=]{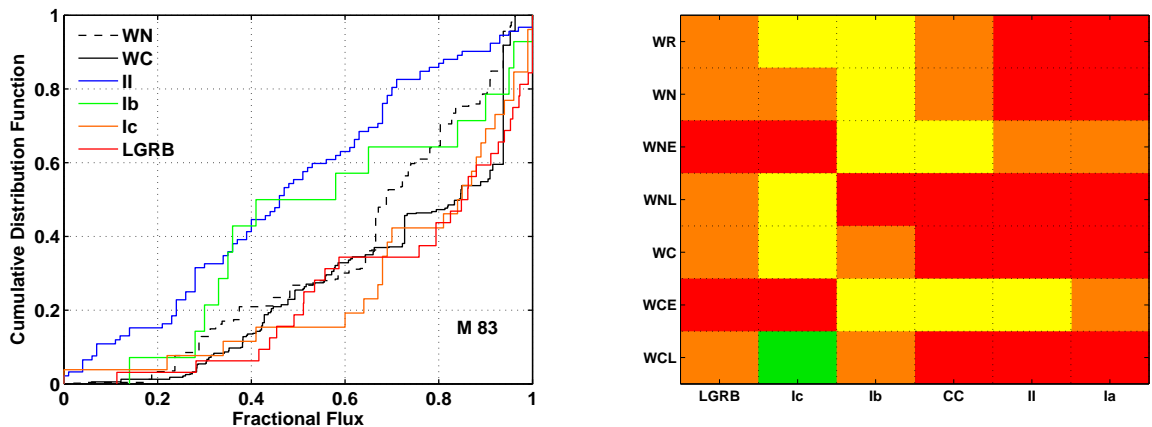}
\caption{
{\bf Left panel:} The distribution of WN and WC stars in M~83, a galaxy with properties typical of the K08 SN Ib/c sample, 
with respect to their location on their host light, plotted with the 
distributions of LGRBs and SNe~II, Ib, and Ic (F06, K08).
{\bf Right panel:}  This colour map indicates at what significance level we can exclude that a SN or LGRB explosion is associated to a certain (sub)type of a progenitor WR star.
Red shows p-values $<0.3\%$, orange $0.3 < p < 4.6\%$, yellow $4.6 < p < 31.7\%$, and green $p > 31.7\%$ (Table~\ref{tab:KSstarnumber}).
The comparison has been done after the removal of the bulge, because
no detailed information about the nuclear WR population exists and a bulge can be visually identified and removed (K08).
The WC distribution is more skewed towards brighter pixels than the WN distribution. 
It is more probable that WC stars are drawn from the same population as SNe~Ic
than SNe~Ib (or any other type of SN), and it is more probable that WN stars are associated to SNe~Ib than with SNe~Ic (or any other type of SN).
This is consistent with theoretical predictions \citep{WRreview,Georgy2009}.
}
\label{fig:M83cdfnKS}
\end{figure*}

The pixel fractional fluxes measured at the locations of the WR stars,
together with their number and subtype, are listed for the two
galaxies in Table~\ref{tab:FFpart}.   In Figs.~\ref{fig:M83cdfnKS} and \ref{fig:NGC1313cdfnKS} (left panels), we
have plotted the WN and WC fractional flux distributions for the two galaxies, 
together with the data from
F06 and K08.  Table~\ref{tab:KSstarnumber} contains the corresponding
Kolmogorov-Smirnov (KS) test
p-values that two distributions are drawn from the same parent
distribution. To keep the graph as uncrowded as possible, we
focused on the most relevant WR and SN types.  The table,
however, contains more information including the division of WR stars
into their subtypes.

The null hypothesis of the KS test is that two samples
are drawn from the same distribution and the purpose of the test is to
reject (or not) this null hypothesis. 
The KS test should not be used to deduce new physics, 
but rather 
for the opposite purpose, namely to test whether 
an idea with strong theoretical background, such as that WR stars
and some  CC SNe are associated, can be rejected.
In that respect, a definite rejection of the null hypothesis requires a p-value $<0.3\%$.
Doubts can exist for p-values $<5\%$, but these are certainly not enough to disprove a well-justified hypothesis.
Inspired by the Gaussian distribution, 
from now on 
we call these significance levels 
(or rather the very similar 0.3\% and 4.6\%)
the $3 \sigma$ and 2$\sigma$ levels at which the hypothesis can be rejected, although this is purely a naming convention. 
For higher p-values, including the (Gaussian) $1 \sigma$, or 31.7\% limit, there is very weak evidence against the null hypothesis.

To better illustrate this, in Figs.~\ref{fig:M83cdfnKS} and \ref{fig:NGC1313cdfnKS} (right panels) 
we have colour-mapped the
KS p-values from Table~\ref{tab:KSstarnumber} reflecting their significance levels.
The only conclusive result ($>3 \sigma$ exclusion) is coloured in red. Orange, yellow, and green show the progressive decreasing 
significance at which the hypothesis of common parent distribution can be rejected. 
In that context, yellow is  more probable than orange, but even orange
cannot be excluded by the present data.
From now on, when we refer to our `results', we are mostly refering to these significance levels and their relative order.

\begin{table}
\caption{KS test p-values (\%).}
\label{tab:KSstarnumber}
\centering
\begin{tabular}{llrrrrrrr}
\hline\hline
 & 	& LGRB &      Ic &      Ib  &	 CC  &     II	&    Ia  \\ 
\hline
 & WR	& 1.2 &    20.7 &     8.3  &	2.1  &    0.0	&   0.0  \\ 
 & WN	& 0.7 &     2.4 &    16.6  &	1.8  &    0.0	&   0.0  \\ 
 {\bf M~83 } & WNE  & 0.0 &     0.0 &     5.2  &   29.4  &    1.5	&   1.5  \\ 
{\bf (High {\it Z})}  & WNL  & 3.5 &    29.8 &     0.2  &	0.0  &    0.0	&   0.0  \\ 
 & WC	& 2.0 &    31.1 &     4.6  &	0.2  &    0.0	&   0.0  \\ 
 & WCE  & 0.0 &     0.0 &     8.3  &   11.5  &    5.0	&   3.7  \\ 
 & WCL  & 2.2 & 	33.0 &     2.6  &    0.1  &    0.0   &   0.0  \\    
 \hline  
 & WR	&     0.3   &   0.2  &   77.2  &   53.0   &   7.5   &   9.4  \\ 
 & WN	&     0.1   &   0.0  &   81.3  &   96.4   &  68.1   &  89.7  \\ 
 {\bf NGC~1313} & WNE  &     5.9   &   8.1  &   47.6  &   64.4   &   6.5   &  10.3  \\ 
 {\bf (Low {\it Z})} & WNL  &     0.0   &   0.0  &   22.9  &   25.1   &  20.0   &  15.3  \\ 
 & WC	&    17.9   &  14.5  &   22.8  &    6.1   &   1.0   &   0.8  \\ 
 & WCE  &    17.9   &  14.5  &   22.8  &    6.1   &   1.0   &   0.8  \\
  &  WCL 	 &    \nodata   &   \nodata   &   \nodata   &   \nodata  &    \nodata  & 	\nodata  \\
   \hline
\end{tabular} \\
\end{table}

\subsection{High metallicity -- M~83}

For the metal-rich galaxy M~83, which is typical of the K08 SNe Ib/c 
sample, we make the following observations. 

\labelitemiii~
The distribution of WR stars as a whole is consistent with those of SNe Ib/c (yellow in Fig.~\ref{fig:M83cdfnKS}) or even LGRBs (orange).
It is however inconsistent with SNe II (red).

\labelitemiii~
The distribution of WC stars is more consistent with SNe~Ic 
($p=31.5\%$, almost green) than with any other type of SN ($p<4.6\%$, which occurs for SNe Ib).
From the WC subtypes, it is mostly the WCL stars that are
responsible for this association.
The less abundant WCE stars, on the other hand, seem to better follow the host galaxy light and
are more consistent with SNe that behave in a similar way.

\labelitemiii~ 
WN stars are more consistent with SNe~Ib (yellow) than
with other kinds of supernovae (orange or red).
We caution, however, that the early and late WN distributions behave quite differently. 
By excluding WNL stars, which might not be direct SN progenitors 
(see Sect.~\ref{disc}), 
WNE stars alone show a clearer preference to SNe~Ib and 
their association to SNe~Ic can be ruled out with a certainty of  over 3$\sigma$.

\labelitemiii~ 
SNe~II and Ia, which follow their host galaxy light distribution well (K08), 
show no association (null hypothesis excluded at over 3$\sigma$) 
with most WR stars, with the possible exception of early subtypes that seem to occupy fainter locations than their late counterparts.
The same claim, but not as strict, could be made for the high-redshift CC SN sample of F06 
(which most likely consists of SNe~II).

To study how the errors in the number of WR stars can affect the
results above, we have followed an MC approach.  Multiple realizations
of the distributions were generated where the numbers of WR stars per
region were drawn randomly from Gaussian distributions with the mean
and standard deviation specified by the number of WR stars and the associated 
errors provided by \cite{M83}.  While this causes the p-values to fluctuate
around their central values in Table~\ref{tab:KSstarnumber}, none of
the qualitative conclusions above are affected.  
By this, we mean that p-values rarely jump to another $\sigma$ significance level 
(i.e. their colour in Fig.~\ref{fig:M83cdfnKS} does not change) 
and that their relative ordering remains the same.  
The standard deviations on the p-values scale with the p-values themselves:
typical fluctuations are of the order of  $\pm$6\% for a p-value of 26\%  or $\pm$1\% for a p-value of 3.4\%.  
Of course, if the number estimates of \cite{M83}
are biased in a  systematic way, significant changes might be expected.  
The results may also be susceptible to changes in the limited number of
SNe and LGRBs in the comparison samples.

The above results were obtained after removing the 
bulge contribution of M~83 and comparing with the corresponding results of K08.
This is because we have no detailed information about the nuclear WR population of this galaxy, 
and a bulge can be visually identified and removed. 
The bulge light removal was done, similar to K08, by placing a circular ring around the bulge 
and by replacing all bulge pixels with the mean pixel value inside the ring.

\subsection{Low metallicity -- NGC~1313}
\label{res1313}

In the case of the metal-poor NGC~1313,  we observe the following:

\labelitemiii~
The global WR population is this time mostly consistent with SNe~Ib 
(at a highly significant p-value), while at the same time 
an association with H-rich SNe is probable as well.

\labelitemiii~
This is especially obvious in the case of WN stars that 
trace well the light of the host galaxy. 
WN stars are again mostly consistent with SNe~Ib, while a relation to SNe~Ic can be excluded this time at 
high significance.  
On the other hand, their association to SNe~II cannot be excluded any more, but instead shows a high probability.
Again, differences are seen between the WNE and WNL distributions.

\labelitemiii~
WC stars, on the other hand, show significant association probabilities with SNe~Ic and even LGRBs.
We recall here that the WC population at the metallicity of NGC~1313 
consists entirely of WCE stars.

Again, we checked on how the total WR content in the range
discussed by \cite{NGC1313} affects the distributions and we did not
find any qualitative difference with the results above.  
In the case of NGC~1313, the numbers of WR stars per region are
considerably lower than in M~83.
If region \#64 contains 6 rather than 3 WC stars, as suspected by \cite{NGC1313},
the WC distribution is pushed even closer to the SNe~Ic and further from the SNe~II (which turn red).

\begin{figure*}
\includegraphics[width=\textwidth,clip=]{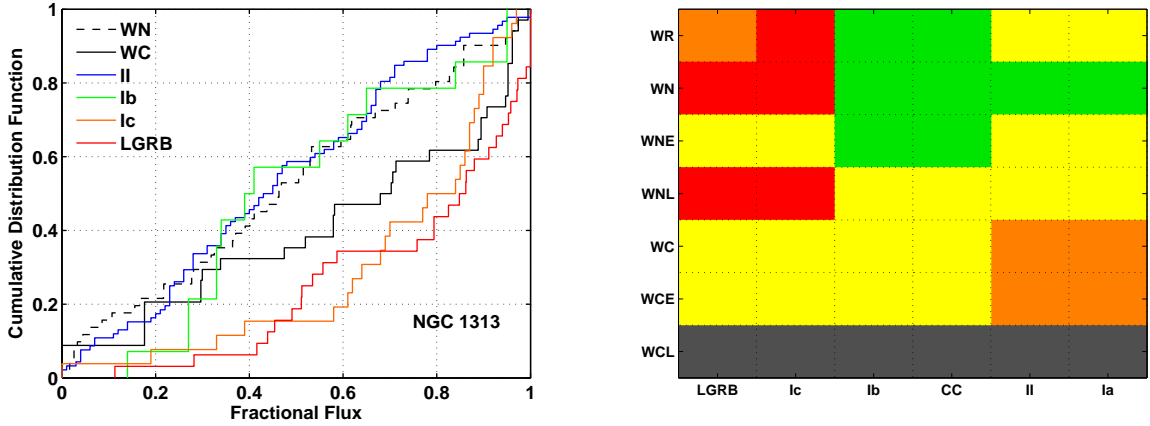}
\caption{
The same as in Fig.~\ref{fig:M83cdfnKS} but for NGC~1313, 
a more metal-poor and irregular galaxy, 
probably similar to
a low-redshift counterpart of the high-redshift F06 hosts.
The colour-coding in the right panel is the same as 
in Fig.~\ref{fig:M83cdfnKS}, but grey has been used for WCL stars that are not present in this galaxy.
}
\label{fig:NGC1313cdfnKS}
\end{figure*}

\section{Discussion}
\label{disc}

From the results presented in the previous section, a qualitative
pattern seems to emerge: WC stars are on average found in
brighter pixels than WN stars. 
As a consequence, WC stars show higher
probabilities of association with SNe~Ic.  On the other hand, WN
stars are most consistent with the locations of SN~Ib explosions.
This is the main result presented in this paper 
and is in broad agreement  with the theoretical
expectations, i.e., that SNe~Ic result from progenitors that have 
been stripped of a larger part of their outer envelope \citep{WRreview}.  
Also from a statistical point of view the WN and WC distributions are almost (geographically) incompatible 
(at significance $>3 \sigma$ for M~83, while $p=5.9\%$ for NGC~1313), which highlights the need to consider these two subtypes separately
when discussing SN progenitors, despite their strong physical connection.
Other studies, however,
caution that there is no strict one-to-one
correlation between progenitor and supernova type,
but that leaks might exist, e.g., less massive WC stars exploding as SNe~Ib \citep{Georgy2009}.

Concerning LGRBs, they are not inconsistent with being drawn from the WR population, although typically at lower
significance than SNe~Ic.
This is not a surprise since not 
all SNe Ic produce LGRBs \citep[e.g.][]{soderbergRadioIbc}.
It is, however, tempting to point out that the highest p-value obtained for LGRBs (yellow) is the one for WC stars at low metallicity,
in agreement with the proposed low-metallicity requirement \citep{YoonLanger,WoosleyHeger}.

Besides the general trend that is common for the two galaxies, differences do exist between 
the individual WN and WC fractional flux distributions. 
The difference is more pronounced in the case of the WN distributions that are almost mutually inconsistent ($p=0.6\%$). 
The WN distribution in NGC~1313 tracks the host galaxy light better and is  more consistent with SNe~II. 
To some degree, this difference can be attributed to the important metallicity difference between the two galaxies:  
\cite{Georgy2009} predict that at low metallicity (similar to NGC~1313) the highest fraction of WN
stars are actually expected to explode as SNe~II and not SNe~Ib. 
Due to lower mass-loss, the less massive stars are still expected to leave 
enough hydrogen to be detectable in the explosion spectra.

Other differences, however, especially at the subtype level, are more 
difficult to explain.  The most striking is related
to the WNL populations of the two galaxies. In M~83 they are found on
the brightest pixels, while in NGC~1313 they lie on the faintest ones. 
Indeed, the two WNL distributions are
inconsistent with each other, at a significance $>3 \sigma$.
The reason for this is unclear, but it
should be mentioned that WNL stars are not always
stripped-envelope massive stars.  
In many cases, they are very luminous
H-rich WN stars that are still burning H in their core. They are
therefore in a phase preceding the LBV phase and
\emph{not} direct progenitors of SNe Ib/c \citep{WRreview,WNH}.
According to \cite{WRreview}, a possible way to distinguish between
H-rich WNL stars and stripped WNL stars is that the former usually lie
in young massive clusters.  One could argue that this is mostly the
case for the WNL stars in M~83, while the ones in NGC~1313 are mostly isolated.  
If this is true, caution should be applied when comparing the WNL distribution of M~83 to those of SNe or LGRBs.
(WNE stars alone share most of the qualitative properties discussed for WN stars in this galaxy.)
On the other hand, 
their positions in NGC~1313 are consistent with the predictions of 
\cite{Georgy2009}: 
that WNL stars are expected to give SNe~II, especially at low
metallicity.  
We may thus be probing different WNL populations in the two galaxies.

Below we assess the robustness of our
results, with respect to several uncertainty factors.  
We call a result robust as long as the significance levels 
in the KS test between two distributions 
(in the form of colours in Figs.~\ref{fig:M83cdfnKS} and~\ref{fig:NGC1313cdfnKS}) 
remain unchanged and retain their relative values with respect to other p-values.
Indeed, with small exceptions,  this is the case for most p-values.
We conclude therefore that our main results are not sensitive to these uncertainties.

\subsection{The remaining candidates' locations and their implications.}
\label{cand_loc}

The number and nature of the remaining candidate regions in our galaxies has until now been ignored.
To determine how important that is, we have followed an MC approach.
By comparing the fractional flux distribution of the candidate pixels of M~83 to the corresponding distribution of the WN and WC pixel positions
(i.e. without taking the number of stars per region into account this time, since we lack this kind of information for the candidate regions)
we find that the candidate locations are more likely drawn from the WN ($p=60\%$)  than the WC ($p=0.5\%$) pool. 
A possible reason could be that WC stars have stronger narrow-band excess over the continuum and might have been 
preferentially selected for 
follow-up spectroscopy.
Indeed, WN stars, especially weak-lined WNE and very late WNL types, might suffer from some incompleteness (P.~Crowther, priv. comm.).
For NGC~1313, \cite{NGC1313} state that the (few) remaining candidate regions are photometrically 
consistent with WN stars.
 
\begin{figure}
\includegraphics[width=\columnwidth,clip=]{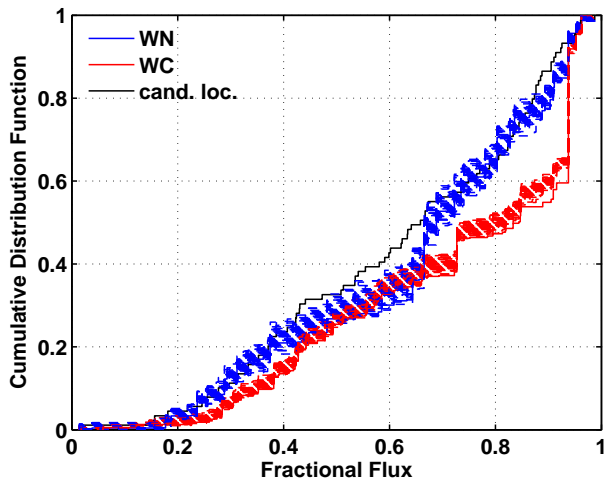}
\caption{Example MC simulation for the inclusion of 
the candidate regions in the WR distributions in M~83. 
Each realization results in a different WN and WC distribution (dashed lines), 
while the solid lines denote the original distributions (Fig.~\ref{fig:M83cdfnKS}).
For visual purposes, only the first 50 realizations have been plotted although the simulation contains 1000 runs.
In the displayed simulation, the candidates were divided to WC and WN stars according to the observed WC/WN ratio.
Our main conclusions remain unchanged, as do the colours in Fig.~\ref{fig:M83cdfnKS}, right panel.
Only the p-value between WC stars to SNe~Ib changed colour, becoming yellow ($p=6.6 \pm 1.3$\%) from being (marginally) orange ($p=4.6$\%).
Similar MC simulations were used for both galaxies in order to assess the importance of effects like 
the presence of foreground stars fainter than 20 mag, 
that the number of stars per region have an associated error estimate,
and the evolution of WN to WC stars.
}
\label{fig:MCcandM83.eps}
\end{figure}

In our MC simulation for M~83, we attempted many realizations where we allowed 66\% of the candidates
(a percentage equal to the success rate of the actual spectroscopic survey) to be genuine WR regions
containing a number of stars equal to the median number of stars per confirmed WR region (plus their median error). 
We made simulations for the two limiting cases that the candidate regions included in the actual WR distributions 
were all included in the WN distribution or
were divided between WN and WC stars according to the observed WC/WN ratio.
The latter simulation is displayed as an example in Fig.~\ref{fig:MCcandM83.eps}.
In both simulations we see only small differences in the results reported 
in Table~\ref{tab:KSstarnumber}, which do not change any of our conclusions.
Similarly, we found no significant differences in our main results for NGC~1313.

\subsection{The effect of temporal evolution}
\label{sec:evol}

A possible objection concerning the significance between the apparent difference between the WN and WC distributions
is that sometimes a WN star is just on an evolutionary path towards a WC final stage, while all WC stars have already been through the WN phase once \citep{MeyMae2005,WRreview}.
Whether this evolution will occur at all, and the related timescale, 
is strongly dependent on mass and metallicity \citep{MeyMae2005}.
The mass range for which stars actually die as WN stars is quite narrow \citep{Georgy2009} but if  convolved with the IMF
 their number can become important.
 By using the mass limits from \citeauthor{Georgy2009} (\citeyear{Georgy2009}; their table 4), we estimate that, at the metallicity of NGC~1313, only $\sim1/6$ of the 
 observed WN stars (the most massive) will evolve to WC stars. In M~83, however, we expect this ratio to increase to almost $2/3$.

In an MC simulation for each galaxy, we allowed the above-mentioned ratios of WN stars to be removed from the WN and be included
in the WC distributions.
The expected result is  that the WC distribution will be pushed towards fainter pixels, since it will be contaminated by WN
stars.
This simulation ignores 
that, while some WN stars evolve to WC, 
some new WN stars will be born, and many stars will explode. 
It also ignores the related timescales (all of a few Myr), and
the brightness evolution of the stars themselves and has deliberately 
not made any 
assumptions about masses based on the positional information of the stars on the galaxy.
A hidden (reasonable) assumption is therefore that we do not observe these galaxies at a very special time 
in their existence 
and that, in this context, these simulations represent a limiting worst case.
For both galaxies, we note that the 
probability of associating
WC to SNe~Ib jumps up one significance level, while their
association to SNe~Ic and LGRBs is reduced, although the significance remains the same.
The WN central association p-values remain unchanged, as expected, while all WN and WC p-values get assigned error bars 
that can occasionally cross different significance levels. 
However, even in this  limiting worst case, the relative scaling between the p-values remains unchanged and the WC distribution
is always skewed towards brighter pixels than WN.

\subsection{How much does binning affect the results?}

To compare with results obtained for galaxies at higher redshift, our images were subject to the degradation process (binning)
described in Sect.~\ref{sec:pre-proc}. 
We examine here the implications of this process.

If the fractional fluxes are measured in the original images, 
considerably higher values are obtained and the WR distributions become skewed towards brighter pixels,
even brighter than for distant 
SNe~Ic and LGRBs, especially in the case of WC stars.
The effect of binning is that a bright isolated star 
(that has a high pixel fractional flux value in our original image) 
will be smoothed out  and have a low fractional flux in the processed image. 
A bright association of many pixels, however, such as a cluster, will be less affected and therefore 
have a high fractional flux also in the processed image.
This mimics the effect of distance, where isolated stars cannot be detected at higher redshift  but clusters can.
Although not binning would result in an apparently stronger result with regard to the probable association 
of WR stars to SNe~Ib/c  and LGRBs,  
this degradation is needed to make a fair comparison. 
In that context, if WR stars are indeed the progenitors of these explosions, the fractional flux values at the low brightness tails of the F06 and K08
distributions, are probably caused by isolated WR stars, while the ones with high fractional fluxes are those that are found in bright clusters.
That these explosions tend to occur in pixels brighter than average (F06, K08) can then be explained 
by the preference of WR stars to be found in large associations. 

We have also tested various degrees of binning to confirm 
that our results are not tuned to the chosen values (simulated redshift $\sim$0.01).
Although some small changes occur, 
we checked that our conclusions are robust to lower and higher 
values of binning as long as the  
PSF in the original image is not subsampled.

\section{Conclusions}
\label{conc}

We performed a F06 type analysis on the WR populations 
of two nearby galaxies 
and  compared our results to the distributions of different
types of SNe (K08), focusing on SNe Ib/c, and LGRBs (F06).  
M~83 is a metal-rich galaxy, typical of the K08 SNe Ib/c host sample, while
NGC~1313 is more metal poor,  irregular, and similar to high-redshift LGRB
host galaxies.  To enable the comparison we resampled our images to simulate a higher redshift.

WR stars are consistent with being the progenitors of SNe~Ib/c or even LGRBs.
Furthermore, the WC stars are distributed in brighter locations of their hosts than WN stars.
It is therefore more likely that WC stars are the progenitors of SNe~Ic and WN stars of SNe~Ib, as also expected by theoretical arguments. 
This result is robust to a number of systematic checks that we carried out.

Although encouraging, these results are based on the only two galaxies for which such an analysis is possible at present. 
Even though they contain enough of WR stars and we have shown that they are, most probably, not special in any way, 
it would of course be desirable to validate our results on a larger galaxy sample, once such a sample becomes available. 
Ideally, such a sample should span a wide range of metallicities.

\begin{acknowledgements}
We thank Paul Crowther for reading the manuscript and giving us comments, as well as 
Jens Hjorth and Micha\l\ Micha\l owski for discussions.
We are grateful to Davide De Martin and to the ESO ePOD department for providing us with a reduced image of M~83 from the ESO/MPI 2.2m.
The Dark Cosmology Centre is funded by the Danish National Research Foundation. 
JS is a Royal Swedish Academy of Sciences Research Fellow supported by a grant from the Knut and Alice Wallenberg Foundation. 
JRM is a Sophie \& Tycho Brahe Fellow.
This research has made use of the NASA/IPAC Extragalactic Database (NED), which is operated by the Jet Propulsion Laboratory, California Institute of Technology, under contract with the National Aeronautics and Space Administration.
Funding for the SDSS and SDSS-II has been provided by the Alfred P. Sloan Foundation, the Participating Institutions, the National Science Foundation, the U.S. Department of Energy, the National Aeronautics and Space Administration, the Japanese Monbukagakusho, the Max Planck Society, and the Higher Education Funding Council for England. 

\end{acknowledgements}


\bibliographystyle{aa}  
\bibliography{13753.bib}

\clearpage

\begin{appendix}
\section{Online Table}

\longtab{1}{

\begin{longtable}{lrrrr}
\caption{\label{tab:FFwhole} Fractional fluxes at the locations of the WR stars. Full version of Table~\ref{tab:FFpart} for the online article.}\\
\hline\hline
& ID & Subtype  &   Number & FF     \\
\hline
\endfirsthead
\caption{continued.}\\
\hline\hline
& ID & Subtype  &   Number & FF    \\
\hline
\endhead
\hline
\endfoot
{\bf M~83}   &  	 1   &    WNL	&	3  &   0.29    \\  
  &  	 2   &    WCL	&	2  &   0.34    \\  
  &  	 3   &    WCL	&	2  &   0.26    \\  
  &  	 4   &    WNE	&      16  &   0.48    \\  
  &  	 5   &    WCL	&	8  &   0.48    \\  
  &  	 6   &    WCE	&	3  &   0.06    \\  
  &  	 7   &    WNL	&	5  &   0.61    \\  
  &  	 8   &    WCL	&	1  &   0.40    \\  
  &  	 9   &    WCE	&	1  &   0.58    \\  
  &  	10   &    WCL	&	2  &   0.54    \\  
  &  	11   &    WNE	&	4  &   0.26    \\  
  &  	12   &    WNE	&	2  &   0.36    \\  
  &  	13   &    WCE	&	3  &   0.56    \\  
  &  	14   &    WCL	&	4  &   0.12    \\  
  &  	15   &    WCE	&	1  &   0.38    \\  
  &  	16   &    WCE	&	2  &   0.44    \\  
  &  	17   &    WCL	&	6  &   0.42    \\  
  &  	18   &    WCL	&	6  &   0.38    \\  
  &  	19   &    WCL	&	1  &   0.38    \\  
  &  	20   &    WCL	&	2  &   0.35    \\  
  &  	21   &    WCL	&	1  &   0.30    \\  
  &  	22   &    WNE	&	4  &   0.34    \\  
  &  	23   &    WCE	&	1  &   0.29    \\  
  &  	24   &    WCL	&	4  &   0.28    \\  
  &  	25   &    WNL	&	4  &   0.32    \\  
  &  	26   &    WNL	&	2  &   0.72    \\  
  &  	27   &    WCL	&	6  &   0.85    \\  
  &  	28   &    WCL	&	8  &   0.85    \\  
  &  	29   &    WCL	&	1  &   0.55    \\  
  &  	30   &    WCL	&	1  &   0.23    \\  
  &  	31   &    WCL	&      42  &   0.73    \\  
  &  	32   &    WCL	&      12  &   0.43    \\  
  &  	33   &    WNL	&	1  &   0.71    \\  
  &  	34   &    WCL	&	3  &   0.51    \\  
  &  	35   &    WNE	&      14  &   0.72    \\  
  &  	36   &    WCL	&	5  &   0.72    \\  
  &  	37   &    WCL	&	4  &   0.91    \\  
  &  	38   &    WNL	&	7  &   0.95    \\  
  &  	38   &    WCL	&      21  &   0.95    \\  
  &  	39   &    WCE	&	2  &   0.81    \\  
  &  	40   &    WCL	&	6  &   0.89    \\  
  &  	41   &    WNL	&      14  &   0.91    \\  
  &  	41   &    WCL	&      13  &   0.91    \\  
  &  	42   &    WNL	&	5  &   0.95    \\  
  &  	43   &    WNL	&	3  &   0.86    \\  
  &  	44   &    WNL	&	5  &   0.74    \\  
  &  	45   &    WNL	&	1  &   0.63    \\  
  &  	46   &    WCL	&	3  &   0.82    \\  
  &  	47   &    WCE	&	1  &   0.69    \\  
  &  	48   &    WNE	&      44  &   0.67    \\  
  &  	49   &    WCL	&      13  &   0.49    \\  
  &  	50   &    WNL	&	8  &   0.69    \\  
  &  	51   &    WNL	&	1  &   0.23    \\  
  &  	52   &    WNE	&      18  &   0.29    \\  
  &  	53   &    WCL	&	2  &   0.23    \\  
  &  	54   &    WCL	&	6  &   0.56    \\  
  &  	55   &    WNE	&      20  &   0.24    \\  
  &  	56   &    WNE	&      15  &   0.69    \\  
  &  	57   &    WNL	&	9  &   0.73    \\  
  &  	58   &    WCL	&	5  &   0.42    \\  
  &  	59   &    WNL	&      14  &   0.83    \\  
{\bf M~83}   &  	60   &    WNE	&	6  &   0.75    \\  
  &  	61   &    WCL	&	2  &   0.84    \\  
  &  	62   &    WCL	&	4  &   0.44    \\  
  &  	63   &    WCL	&	5  &   0.28    \\  
  &  	64   &    WNL	&      10  &   0.31    \\  
  &  	65   &    WCL	&	4  &   0.52    \\  
  &  	66   &    WNL	&	8  &   0.59    \\  
  &  	66   &    WCL	&	4  &   0.59    \\  
  &  	67   &    WCL	&	5  &   0.57    \\  
  &  	68   &    WCL	&	3  &   0.44    \\  
  &  	69   &    WCL	&	2  &   0.28    \\  
  &  	70   &    WCL	&	2  &   0.25    \\  
  &  	71   &    WCL	&	3  &   0.62    \\  
  &  	72   &    WCL	&	1  &   0.75    \\  
  &  	73   &    WNL	&	2  &   0.74    \\  
  &  	74   &    WNL	&      52  &   0.94    \\  
  &  	74   &    WCL	&     179  &   0.94    \\  
  &  	75   &    WCL	&	1  &   0.39    \\  
  &  	76   &    WNL	&	1  &   0.66    \\  
  &  	77   &    WCL	&	2  &   0.61    \\  
  &  	78   &    WNL	&      11  &   0.91    \\  
  &  	79   &    WCL	&	5  &   0.66    \\  
  &  	80   &    WNL	&	1  &   0.33    \\  
  &  	81   &    WCL	&	3  &   0.46    \\  
  &  	82   &    WCL	&	4  &   0.61    \\  
  &  	83   &    WCL	&	2  &   0.56    \\  
  &  	84   &    WCL	&	9  &   0.92    \\  
  &  	85   &    WCL	&	4  &   0.58    \\  
  &  	86   &    WNL	&	9  &   0.96    \\  
  &  	86   &    WCL	&      24  &   0.96    \\  
  &  	87   &    WCE	&	3  &   0.41    \\  
  &  	88   &    WCL	&	1  &   0.39    \\  
  &  	89   &    WCL	&	3  &   0.79    \\  
  &  	90   &    WCL	&	2  &   0.82    \\  
  &  	91   &    WCL	&	3  &   0.45    \\  
  &  	92   &    WCL	&	3  &   0.31    \\  
  &  	93   &    WNE	&      10  &   0.19    \\  
  &  	94   &    WNE	&      23  &   0.64    \\  
  &  	95   &    WNL	&	1  &   0.43    \\  
  &  	96   &    WNL	&	2  &   0.79    \\  
  &  	97   &    WNL	&	5  &   0.90    \\  
  &  	98   &    WCL	&	4  &   0.38    \\  
  &  	99   &    WNL	&	1  &   0.71    \\  
  &    100   &    WNL	&	5  &   0.89    \\  
  &    101   &    WNE	&      11  &   0.45    \\  
  &    102   &    WNL	&	9  &   0.84    \\  
  &    103   &    WNL	&      29  &   0.80    \\  
  &    104   &    WCL	&	3  &   0.73    \\  
  &    105   &    WNL	&	8  &   0.89    \\  
  &    106   &    WCL	&	6  &   0.85    \\  
  &    107   &    WNE	&      15  &   0.78    \\  
  &    108   &    WNL	&	2  &   0.58    \\  
  &    109   &    WCL	&	7  &   0.83    \\  
  &    110   &    WCL	&	3  &   0.64    \\  
  &    111   &    WCL	&	3  &   0.32    \\  
  &    112   &    WCL	&	2  &   0.78    \\  
  &    113   &    WCL	&	5  &   0.38    \\  
  &    114   &    WNE	&	4  &   0.18    \\  
  &    115   &    WCL	&	4  &   0.58    \\  
  &    116   &    WNL	&      12  &   0.67    \\  
  &    117   &    WNE	&	1  &   0.33    \\  
{\bf M~83}   &    118   &    WCL	&	6  &   0.66    \\  
  &    119   &    WCL	&	2  &   0.51    \\  
  &    120   &    WNE	&	6  &   0.53    \\  
  &    121   &    WNE	&      17  &   0.37    \\  
  &    122   &    WCE	&	3  &   0.27    \\  
  &    123   &    WCL	&	2  &   0.42    \\  
  &    124   &    WCL	&	3  &   0.31    \\  
  &    125   &    WCL	&	3  &   0.31    \\  
  &    126   &    WCE	&	1  &   0.49    \\  
  &    127   &    WCE	&	3  &   0.30    \\  
  &    128   &    WCL	&	2  &   0.39    \\  
  &    129   &    WCE	&	4  &   0.35    \\  
  &    130   &    WNE	&	1  &   0.04    \\  
  &    131   &    WCL	&	1  &   0.27    \\  
  &    132   &    WNE	&	1  &   0.02    \\  
  &    133   &    cand.   &   \nodata  &   0.25      \\  
  &    134   &    cand.   &   \nodata  &   0.54      \\  
  &    135   &    cand.   &   \nodata  &   0.38      \\  
  &    136   &    cand.   &   \nodata  &   0.41      \\  
  &    137   &    cand.   &   \nodata  &   0.37      \\  
  &    138   &    cand.   &   \nodata  &   0.43      \\  
  &    139   &    cand.   &   \nodata  &   0.60      \\  
  &    140   &    cand.   &   \nodata  &   0.62      \\  
  &    141   &    cand.   &   \nodata  &   0.18      \\  
  &    142   &    cand.   &   \nodata  &   0.31      \\  
  &    143   &    cand.   &   \nodata  &   0.27      \\  
  &    144   &    cand.   &   \nodata  &   0.82      \\  
  &    145   &    cand.   &   \nodata  &   0.73      \\  
  &    146   &    cand.   &   \nodata  &   0.75      \\  
  &    147   &    cand.   &   \nodata  &   0.51      \\  
  &    148   &    cand.   &   \nodata  &   0.95      \\  
  &    149   &    cand.   &   \nodata  &   0.91      \\  
  &    150   &    cand.   &   \nodata  &   0.95      \\  
  &    151   &    cand.   &   \nodata  &   0.95      \\  
  &    152   &    cand.   &   \nodata  &   0.87      \\  
  &    153   &    cand.   &   \nodata  &   0.86      \\  
  &    154   &    cand.   &   \nodata  &   0.79      \\  
  &    155   &    cand.   &   \nodata  &   0.31      \\  
  &    156   &    cand.   &   \nodata  &   0.89      \\  
  &    157   &    cand.   &   \nodata  &   0.76      \\  
  &    158   &    cand.   &   \nodata  &   0.42      \\  
  &    159   &    cand.   &   \nodata  &   0.91      \\  
  &    160   &    cand.   &   \nodata  &   0.96      \\  
  &    161   &    cand.   &   \nodata  &   0.30      \\  
  &    162   &    cand.   &   \nodata  &   0.53      \\  
  &    163   &    cand.   &   \nodata  &   0.82      \\  
  &    164   &    cand.   &   \nodata  &   0.25      \\  
  &    165   &    cand.   &   \nodata  &   0.59      \\  
  &    166   &    cand.   &   \nodata  &   0.28      \\  
  &    167   &    cand.   &   \nodata  &   0.16      \\  
  &    168   &    cand.   &   \nodata  &   0.86      \\  
  &    169   &    cand.   &   \nodata  &   0.33      \\  
  &    170   &    cand.   &   \nodata  &   0.87      \\  
  &    171   &    cand.   &   \nodata  &   0.82      \\  
  &    172   &    cand.   &   \nodata  &   0.35      \\  
  &    173   &    cand.   &   \nodata  &   0.81      \\  
  &    174   &    cand.   &   \nodata  &   0.37      \\  
  &    175   &    cand.   &   \nodata  &   0.83      \\  
  &    176   &    cand.   &   \nodata  &   0.58      \\  
  &    177   &    cand.   &   \nodata  &   0.74      \\  
  &    178   &    cand.   &   \nodata  &   0.74      \\  
{\bf M~83}   &    179   &    cand.   &   \nodata  &   0.99      \\  
  &    180   &    cand.   &   \nodata  &   0.43      \\  
  &    181   &    cand.   &   \nodata  &   0.15      \\  
  &    182   &    cand.   &   \nodata  &   0.42      \\  
  &    183   &    cand.   &   \nodata  &   0.63      \\  
  &    184   &    cand.   &   \nodata  &   0.63      \\  
  &    185   &    cand.   &   \nodata  &   0.91      \\  
  &    186   &    cand.   &   \nodata  &   0.62      \\  
  &    187   &    cand.   &   \nodata  &   0.60      \\  
  &    188   &    cand.   &   \nodata  &   0.37      \\  
  &    189   &    cand.   &   \nodata  &   0.93      \\  
  &    190   &    cand.   &   \nodata  &   0.92      \\  
  &    191   &    cand.   &   \nodata  &   0.77      \\  
  &    192   &    cand.   &   \nodata  &   0.67      \\  
  &    193   &    cand.   &   \nodata  &   0.34      \\  
  &    194   &    cand.   &   \nodata  &   0.91      \\  
  &    195   &    cand.   &   \nodata  &   0.96      \\  
  &    196   &    cand.   &   \nodata  &   0.88      \\  
  &    197   &    cand.   &   \nodata  &   0.67      \\  
  &    198   &    cand.   &   \nodata  &   0.85      \\  
  &    199   &    cand.   &   \nodata  &   0.67      \\  
  &    200   &    cand.   &   \nodata  &   0.84      \\  
  &    201   &    cand.   &   \nodata  &   0.55      \\  
  &    202   &    cand.   &   \nodata  &   0.33      \\  
  &    203   &    cand.   &   \nodata  &   0.64      \\  
  &    204   &    cand.   &   \nodata  &   0.67      \\  
  &    205   &    cand.   &   \nodata  &   0.84      \\  
  &    206   &    cand.   &   \nodata  &   0.89      \\  
  &    207   &    cand.   &   \nodata  &   0.88      \\  
  &    208   &    cand.   &   \nodata  &   0.87      \\  
  &    209   &    cand.   &   \nodata  &   0.79      \\  
  &    210   &    cand.   &   \nodata  &   0.44      \\  
  &    211   &    cand.   &   \nodata  &   0.66      \\  
  &    212   &    cand.   &   \nodata  &   0.88      \\  
  &    213   &    cand.   &   \nodata  &   0.54      \\  
  &    214   &    cand.   &   \nodata  &   0.82      \\  
  &    215   &    cand.   &   \nodata  &   0.70      \\  
  &    216   &    cand.   &   \nodata  &   0.47      \\  
  &    217   &    cand.   &   \nodata  &   0.22      \\  
  &    218   &    cand.   &   \nodata  &   0.51      \\  
  &    219   &    cand.   &   \nodata  &   0.42      \\  
  &    220   &    cand.   &   \nodata  &   0.39      \\  
  &    221   &    cand.   &   \nodata  &   0.01  	 \\  
  \hline
{\bf NGC~1313}  &  	 1   &    WNE	&	1  &   0.37	 \\  
  &  	 2   &    WCE	&	1  &   0.30	 \\  
  &  	 3   &    WNL	&	1  &   0.06	 \\  
  &  	 7   &    WNL	&	1  &   0.03	 \\  
  &  	 8   &    WNL	&	1  &   0.16	 \\  
  &  	 9   &    WNE	&	1  &   0.28	 \\  
  &  	10   &    WCE	&	2  &   0.30	 \\  
  &  	11   &    WNE$^a$	&	1  &   0.00	 \\  
  &  	11   &    WCE	&	1  &   0.00	 \\  
  &  	12   &    WCE	&	1  &   0.47	 \\  
  &  	13   &    WNL	&	1  &   0.11	 \\  
  &  	14   &    WNE$^a$	&	1  &   0.67	 \\  
  &  	15   &    WCE	&	1  &   0.34	 \\  
  &  	16   &    WCE	&	1  &   0.71	 \\  
  &  	17   &    WNE	&	1  &   0.71	 \\  
  &  	18   &    WNL	&	1  &   0.22	 \\  
  &  	19   &    WNE	&	1  &   0.02	 \\  
  &  	20   &    WCE	&	1  &   0.70	 \\  
{\bf NGC~1313}  &  	21   &    WNL	&	1  &   0.74	 \\  
  &  	22   &    WNL	&	1  &   0.74	 \\  
  &  	23   &    WNE	&	1  &   0.46	 \\  
  &  	24   &    WNL	&	1  &   0.62	 \\  
  &  	25   &    WNL	&	1  &   0.46	 \\  
  &  	28   &    cand.   &   \nodata  &   0.38      \\  
  &  	29   &    WNE	&	1  &   0.39	 \\  
  &  	30   &    WNL	&	1  &   0.03	 \\  
  &  	31   &    WCE	&	1  &   0.78	 \\  
  &  	32   &    WNL	&	1  &   0.03	 \\  
  &  	33   &    WNE	&	3  &   0.86	 \\  
  &  	34   &    cand.   &   \nodata  &   0.91      \\  
  &  	35   &    WCE	&	1  &   0.89	 \\  
  &  	37   &    WCE	&	1  &   0.99	 \\  
  &  	39   &    WNL	&	1  &   0.52	 \\  
  &  	40   &    WNE	&	1  &   0.84	 \\  
  &  	41   &    WCE	&	1  &   0.71	 \\  
  &  	42   &    WCE	&	1  &   0.68	 \\  
  &  	44   &    WNE	&	1  &   0.53	 \\  
  &  	45   &    WNE	&	1  &   0.41	 \\  
  &  	46   &    WNE	&	1  &   0.62	 \\  
  &  	47   &    WNL	&	1  &   0.62	 \\  
  &  	48   &    WCE	&	2  &   0.58	 \\  
  &  	49   &    WCE	&	1  &   0.58	 \\  
  &  	50   &    WCE	&	1  &   0.89	 \\  
  &  	51   &    WNE	&	1  &   0.83	 \\  
  &  	52   &    WNE	&	1  &   0.60	 \\  
  &  	53   &    WNE	&	1  &   0.80	 \\  
  &  	54   &    cand.   &   \nodata  &   0.60      \\  
  &  	55   &    WNE	&	1  &   0.51	 \\  
  &  	56   &    WCE	&	1  &   0.89	  \\  
  &  	57   &    WCE	&	1  &   0.91	  \\  
  &  	58   &    WNE$^a$	&	3  &   0.97	  \\  
  &  	59   &    WNE$^a$	&	1  &   0.95	  \\  
  &  	59   &    WCE	&	1  &   0.95	  \\  
  &  	60   &    WCE	&	1  &   0.97	  \\  
  &  	61   &    WCE	&	3  &   0.18	  \\  
  &  	62   &    WCE	&	1  &   0.18	  \\  
  &  	63   &    cand.   &   \nodata  &   0.56       \\  
  &  	64   &    WCE	&	3  &   0.95	  \\  
  &  	65   &    cand.   &   \nodata  &   0.98       \\  
  &  	66   &    WCE	&	2  &   0.96	  \\  
  &  	67   &    WNE$^a$	&	1  &   0.96	  \\  
  &  	67   &    WCE	&	1  &   0.96	  \\  
  &  	68   &    cand.   &   \nodata  &   0.45       \\  
  &  	70   &    WNE	&	1  &   0.28	  \\  
  &  	71   &    WNE	&	1  &   0.47	  \\  
  &  	72   &    WNL	&	1  &   0.40	  \\  
  &  	73   &    WNL	&	1  &   0.44	  \\  
  &  	75   &    cand.   &   \nodata  &   0.00       \\  
  &  	76   &    WNE	&	1  &   0.22	  \\  
  &  	77   &    WNL	&	1  &   0.36	  \\  
  &  	78   &    WNE	&	1  &   0.30	  \\  
  &  	79   &    WNE	&	1  &   0.04	  \\  
  &  	80   &    cand.   &   \nodata  &   0.54       \\  
  &  	81   &    WCE	&	1  &   0.52	  \\  
  &  	82   &    cand.   &   \nodata  &   0.64       \\  
  &  	83   &    WNL	&	1  &   0.53	  \\  
  &  	85   &    WNL	&	1  &   0.53	  \\  
  &  	86   &    cand.   &   \nodata  &   0.15       \\  
  &  	87   &    cand.   &   \nodata  &   0.20       \\  
{\bf NGC~1313}  &  	88   &    WCE	&	2  &   0.00	  \\  
  &  	90   &    WNL	&	1  &   0.33	  \\  
  &  	91   &    WNL	&	1  &   0.17	  \\  
  &  	92   &    WNL	&	1  &   0.32	  \\  
  &  	94   &    WNL	&	1  &   0.09	  \\  
\end{longtable}
\begin{tabular}{lll}
$^a$ These are stars classified as WN5-6 and we have included them in the WNE distribution.\\
\end{tabular}

}	

\end{appendix}

\end{document}